\documentclass[twocolumn]{aastex701}

\usepackage{graphicx}
\usepackage{caption}
\usepackage{subcaption}
\usepackage{amsmath}

\begin{document}

\title{An Extra-nuclear Water Maser in NGC 1106 Likely Driven By An AGN Shock: Implications for Maser-based Black Hole Mass Measurements}

\author[0000-0002-1292-1451]{Macon Magno}
\affiliation{Department of Physics and Astronomy, Georgia State University, Atlanta, GA 30303, USA}
\affiliation{Department of Physics and Astronomy and George P. and Cynthia Woods Mitchell Institute for Fundamental Physics and Astronomy, Texas A\&M University, 4242 TAMU, College Station, TX 77843-4242}
\affiliation{CSIRO Space and Astronomy, ATNF, PO Box 1130, Bentley WA 6102, Australia}
\email[]{mmagno@gsu.edu}

\author[0000-0003-3168-5922]{Emmanuel Momjian}
\affiliation{National Radio Astronomy Observatory, P.O Box O, Socorro, NM 87801, USA}
\email[]{emomjian@nrao.edu}

\author[0000-0001-5785-7038]{Krista L. Smith}
\affiliation{Department of Physics and Astronomy and George P. and Cynthia Woods Mitchell Institute for Fundamental Physics and Astronomy, Texas A\&M University, 4242 TAMU, College Station, TX 77843-4242}
\email[]{kristalynnesmith@tamu.edu}

\author[0000-0003-4264-3509]{O. Ivy Wong}
\affiliation{CSIRO Space \& Astronomy, PO Box 1130, Bentley, Western Australia 6102, Australia}
\affiliation{ICRAR, University of Western Australia M468, 35 Stirling Highway, Crawley, WA 6009, Australia}
\email[]{Ivy.Wong@csiro.au}

\author[0000-0002-3365-8875]{Travis C. Fischer}
\affiliation{AURA for ESA, Space Telescope Science Institute, 3700 San Martin Drive, Baltimore, MD 21218, USA}
\email[]{tfischer@stsci.edu}

\author[0000-0002-7998-9581]{Michael J. Koss}
\affiliation{Eureka Scientific, 2452 Delmer Street, Suite 100, Oakland, CA 94602-3017, USA}
\email{mike.koss@eurekasci.com}  

\author[0000-0002-4409-3515]{Chenoa D. Tremblay}
\affiliation{SETI Institute, 339 Bernardo Ave, Suite 200, Mountain View, CA 94043, USA}
\affiliation{Berkeley SETI Research Center, University of California, Berkeley, CA 94720, USA}
\affiliation{Department of Physics and Astronomy, University of New Mexico, Albuquerque, NM 87131, USA}
\affiliation{National Radio Astronomy Observatory, 1003 Lopezville Rd., Socorro, NM 87801, USA} 
\email{ctremblay@seti.org}

\correspondingauthor{Macon Magno}
\email{mmagno@gsu.edu}

\begin{abstract}


We present Very Large Array A-configuration ($0.13$\arcsec) 22 GHz spectral-line and continuum observations of the \textit{Swift}-BAT detected Seyfert 2 galaxy, NGC~1106, aimed at localizing a water maser line offset from the host systemic velocity by 541 km s$^{-1}$. The 22 GHz continuum emission reveals a 352 pc extended structure within the nucleus of NGC~1106, and we spatially resolve the water maser emission to a compact region $\sim1$ kpc ($3.12$\arcsec) from the nucleus. Comparing our data with previous 22 GHz 1\arcsec \, observations, we found that the water maser lies near the boundary of a radio extension aligned with the jet axis, likely arising from the interface between the active galactic nucleus (AGN) outflow and the host galaxy's ISM. To verify this, we compared the water maser in NGC~1106 to a known water maser in a star-forming region within the Milky Way, along with known extragalactic water masers associated with star formation and AGN driven outflows. Overall, we found that the maser position, velocity offset, flux density, and alignment with the jet axis are consistent with an origin at the interface between the AGN outflow and the host ISM. Our results have important implications for maser-based black hole mass measurements in which disk geometry is ambiguous. Since single-dish observations cannot spatially separate disk masers from outflow-driven masers, high-velocity maser features attributed to the accretion disk may also include contributions from kpc-scale shocks.

\end{abstract}



\section{Introduction} 

Most galaxies are believed to have a supermassive black hole (SMBH) at their center, and occasionally these SMBHs accrete matter. Actively accreting SMBHs are known as active galactic nuclei (AGN), and have long been known to power outflows and jets of different scales and sizes \citep[e.g.,][]{Perley_2017, Fischer_2021, Sargent_2025, Njeri_2025}, which can profoundly affect their host galaxies through the regulation of star formation \citep[e.g.,][]{Fabian_2012, Smith_2020a, Fischer_2025}. This interaction between AGN-driven outflows and the host galaxy is thought to give rise to the M-$\sigma_{\star}$ correlation and other empirical observations that relate the central black hole to the host galaxy \citep{Gultekin_2009, Boizelle_2021, Cohn_2023}. 

Although AGN can produce powerful relativistic jets, sometimes spanning sizes larger than the host galaxy itself (e.g., Fanaroff-Riley/FR I and II class jets), these represent only a small minority \citep[e.g.,][]{fanaroff_1974, Kellermann_1989, Odea_2009, Fabian_2012, Berton_2020}. In radio-quiet AGN, the primary driver of feedback is more likely to be a radiatively-driven wind originating from the corona and/or accretion disk of the SMBH \citep[e.g.,][]{Kraemer_2020, Fischer_2021, Trindade_2022, Fischer_2023, Fischer_2025, Sargent_2025}; however, radio-quiet AGN have also been known to host small-scale/frustrated radio jets \citep[e.g.,][]{Giroletti_2005, Panessa_2019, Wang_2021, Wang_2025}. All of these types of outflows have been shown to impact their host galaxy in some way, usually through suppressed star-formation rates \citep[e.g., ][]{Smith_2020b}. Inducing star formation and/or depleting the host galaxy of gas usually involves AGN outflows that cause shocks in the interstellar medium (ISM) \citep{Fischer_2019, Panessa_2019, Jarvis_2019, Smith_2020a, Fischer_2025}.

Water masers, most prominently the H$_2$O $6_{16} \rightarrow 5_{23}$ emission line, which occurs at $\lambda = 1.35$ cm / $\nu = 22.235$ GHz, have now been observed in at least 180 galaxies via dedicated surveys, mostly using single-dish radio telescopes\footnote{\url{https://safe.nrao.edu/wiki/bin/view/Main/PublicWaterMaserList}}, such as the Megamaser Cosmology Project \citep[e.g.,][]{Reid_2009, Braatz_2010, Kuo_2011, Kuo_2020}. Water masers are thought to arise primarily from the cold inner regions of the accretion disk around the SMBH and, in this case, they are an extremely valuable dynamical probe of the black hole mass \citep{Maloney_2002a, Greene_2010, Kuo_2011}. Interestingly, water masers have also been frequently observed in Galactic star-forming regions within the Milky Way at the locations of powerful shocks driven by supernovae or stellar winds \citep{Kawamura_1998, Maloney_2002a}. Water masers have also been observed on fairly small scales ($\lesssim 0.4$\arcsec) arising from the interaction between molecular clouds and the radio jet coming from the AGN \citep[e.g.,][]{Gallimore_2001, Xanthopoulos_2001, Peck_2003, Henkel_2005}. If such masers also arise from shocks between AGN-driven outflows and the host galaxy ISM at larger distances ($\sim1$ kpc), then we would have a new lever arm for understanding pc- to kpc-scale AGN feedback. 

NGC~1106 is a \textit{Swift}-BAT detected Compton-thick Seyfert 2 AGN \citep{Baumgartner_2013, Ricci_2015, Koss_2022a, Koss_2022b, Koss_2022c} located at a distance of $63.2$ Mpc \citep[$z = 0.014513 \pm 4.34\times10^{-5}$;][]{Koss_2022b}. The \textit{Swift}-BAT 157-month monthly mosaic light curve shows a fairly steady state in the ultrahard X-rays with a few possible flaring events \citep{Lien_2025}. NGC~1106 is also a known water kilomaser galaxy \citep{Braatz_2008, Tarchi_2011}. The water maser, detected with the Robert C. Byrd Green Bank Telescope (GBT), is a single, narrow (1 km s$^{-1}$) line with a high-velocity offset (523 km s$^{-1}$) relative to the systemic velocity \citep{Braatz_2008}. The single, narrow water maser line was originally thought to come from the accretion disk around the SMBH at the center of NGC~1106 due to its high-velocity offset. Interestingly, \cite{Tarchi_2011} noted that the observed water maser feature in NGC~1106 could be due to star formation or AGN activity; i.e., its origin remains uncertain. Continuum emission from NGC~1106 has been detected in numerous all-sky radio surveys, indicating a fairly compact, unresolved structure down to 2.5\arcsec~angular scales \citep[e.g.,][]{GB6_1, GB6_2, NVSS_1, WENSS, NVSS_2, VLASS}. A recent 22 GHz continuum-imaging campaign unveiled a faint radio extension to the southeast of the compact core \citep{Smith_2016, Smith_2020b, 2025_Magno}. \cite{Shimizu_2017} studied the \textit{Herschel} far-infrared properties of the AGN in the 70-month \textit{Swift}-BAT survey, which contains NGC~1106. They report a dust mass of $10^{7.42^{+0.06}_{-0.07}}$ M$_{\odot}$ with a temperature of $21.12^{+0.98}_{-0.88}$ K and a star-formation rate of $1.15^{+0.17}_{-0.15}$ M$_{\odot}$ year$^{-1}$. \cite{Ichikawa_2019} decomposed the full infrared SED and found a similar infrared luminosity as \cite{Shimizu_2017}, and, therefore, similar infrared properties for NGC~1106.

In this paper, we present spatially resolved observations of the water maser in NGC~1106 and show that it is located $\sim$1~kpc from the nucleus, along the axis of the radio structure, at the boundary of faint radio continuum emission. The observations and data reduction process are presented in Section \ref{sec:obs}, the results in Section \ref{sec:results}, and the discussion in Section \ref{sec:discussion}. In Section \ref{sec:conclusions}, we state our conclusions. Throughout this paper, we assume a cosmology of $H_0 = 69.6$ km s$^{-1}$ Mpc$^{-1}$, $\Omega_M$ = 0.286, and $\Omega_\Lambda = 0.714$ \citep{Bennett_2014}. For all calculations involving redshift, we used \cite{Wright_2006}'s and \cite{2013A&A...558A..33A, 2018AJ....156..123A, 2022ApJ...935..167A}'s Cosmology Calculators. 

\section{Observations and Data Reduction}
\label{sec:obs}

\subsection{Previous Observations}
The water maser in NGC~1106 was targeted with NASA's Deep Space Network (DSN) \citep{Kondratko_2006}. The DSN observations, which did not detect the water maser feature, had a spectral resolution of $3.5$ km s$^{-1}$ and an rms noise of $13$ mJy. These observations were carried out with a $70$-meter single-dish telescope, resulting in an angular resolution of $44$\arcsec~at full width half maximum (FWHM). Observations by \cite{Braatz_2008} were successful in detecting the 74 mJy peak flux density water maser feature within NGC~1106 using the GBT with a spectral setup of $8192$ channels, delivering a channel spacing of $24$ kHz ($0.33$ km s$^{-1}$), and an rms noise of $6$ mJy per channel. The angular resolution achieved in these observations was  $\sim 35$\arcsec~ at FWHM, which is three times smaller than the optical diameter of NGC~1106 \citep{Vaucouleurs_1991}. \cite{Braatz_2008} thought that the detected water maser was originating from the accretion disk based solely on the high-velocity of the water maser.  

\subsection{New VLA Observations}

The observations of the H$_2$O $6_{16} \rightarrow 5_{23}$ maser line at $22.235$ GHz toward NGC~1106 were carried out with the National Science Foundation's Karl G. Jansky Very Large Array (VLA)\footnote{The National Radio Astronomy Observatory (NRAO) is a facility of the National Science Foundation operated under cooperative agreement by the Associated Universities, Inc.} on 2024 December 18 using its K-band receivers (18-26.6\,GHz). The VLA was in its most extended, A configuration, with a maximum baseline length of $36$ km. These observations are part of the project VLA/24B-382.

\begin{deluxetable}{lcc}[t!]
\tablecaption{Parameters for the VLA Observation \label{tab:vla_obs}}
\tablecolumns{3}
\tablehead{ \multicolumn{1}{l}{Parameter} & \multicolumn{1}{l}{Line} & \multicolumn{1}{l}{Continuum} }   
\startdata
Date                                & \multicolumn{2}{c}{2024 December 18} \\
Configuration                       & \multicolumn{2}{c}{A} \\
R.A. of field center (J2000)        & \multicolumn{2}{c}{2$^{\text{h}}$ 50$^{\text{m}}$ 40\fs486} \\
Dec. of field center (J2000)        & \multicolumn{2}{c}{+41\arcdeg 40\arcmin 17\farcs72} \\
Approx. time on source (hrs)        & \multicolumn{2}{c}{2.4} \\
No. of channels                     & $64$                                     & $384$ \\
Total Bandwidth (MHz)               & $128$                                    & $768$ \\
Channel spacing (km s$^{-1}$)       & $27$                                     & $---$\\
Central frequency (GHz)             & $21.960$                                 & $21.9812$\\
FWHM of synthesized beam            & $0$\farcs$131$ $\times$ $0$\farcs$087$   & $0$\farcs$129$  $\times$ $0$\farcs$086$\\
Position Angle of beam              & $76.99$\arcdeg                           & $76.53$\arcdeg \\
RMS noise ($\mu$Jy beam$^{-1}$)     & $176$                                    & $12$\\
\enddata 
\end{deluxetable}

The Wideband Interferometric Digital ARchitecture (WIDAR) correlator was configured to deliver eight $128$ MHz sub-bands with full polarization products along with $64$ spectral channels per sub-band \citep{widar-carlson}. The 2 MHz channel spacing corresponds to $\sim27$ km s$^{-1}$ at the frequency of the 22 GHz water line. We observed 3C~147 as the absolute flux density scale calibrator. The uncertainty in the flux density calibration is expected to be $10-15\%$ \footnote{\url{ https://science.nrao.edu/facilities/vla/docs/manuals/oss/performance/fdscale}}. The source J0251+4315 was observed at 8.5\,GHz to calibrate the pointing of the antennas and at 22\,GHz to calibrate the complex gains. The latter allows for the determination of the absolute position of the target source. The phase-referencing cycle time was $4$ minutes: $3$ minutes on the target and $1$ minute on the complex gain calibrator. The total on-source integration time was 2.4 hours, which yields an RMS noise of 176 $\mu$Jy beam$^{-1}$ for the spectrum and 12 $\mu$Jy beam$^{-1}$ for the continuum spectral windows. All data reduction steps, including flagging, calibration, and imaging, were carried out using the Common Astronomy Software Applications (CASA) version 6.6.4-34 \citep{McMullin_2007, CASA_2020a, CASA_2020b, CASA_2022a, CASA_2022b, TheCASAteam_2022}\footnote{\url{https://casa.nrao.edu/}}. The flagging primarily consists of online system flags, pure zero visibilities, and the channels at the subband edges, in addition to antennas experiencing hardware issues. Table \ref{tab:vla_obs} summarizes the parameters of the observing run. 

We reconstructed the continuum image of NGC~1106 using the \verb|tclean| task in CASA. We interactively cleaned down to a threshold of $55$ $\mu$Jy (\verb|nsigma| = 3) with Briggs weighting (\verb|robust| = 0.8). To image the spectral line data, we followed a similar process as the continuum imaging. We first subtracted the continuum from the data using \verb|uvcontsub|. Then, we set \verb|specmode| to cube, which applies a Doppler Correction to the spectral line data. We also set our spectral frame to be Barycentric (J2000) with optical velocity definition, and our rest frequency is $22.23508$ GHz. The resulting water maser spectrum and the continuum image of NGC~1106 are shown in Figures \ref{fig:water_maser_spectra} and \ref{fig:radio_outflow_image}, respectively. Continuum flux densities were extracted using CASA's \verb|Imaging fitting| dialog, and these values are tabulated in Table~\ref{tab:vla_obs}. 2-D Gaussian fitting was done on the water maser feature using the JMFIT task in the Astronomical Image Processing System \citep[AIPS;][]{Greisen_2003}. 

We have a tentative detection ($5.8\sigma$) of the water maser in NGC 1106. The maser is unresolved with a flux density of $1.02 \pm 0.18$ mJy. The compact nature of the water maser can be seen in Figure~\ref{fig:water_maser_spectra}. Although no new single-dish observations were obtained, we can compare our VLA detection to the earlier GBT measurement by accounting for the difference in spectral resolution. If the VLA line were observed at the $0.3$ km s$^{-1}$ resolution of \citet{Braatz_2008}, its inferred peak flux density would be approximately 85 mJy. Considering the error in the flux density scale is 10-15\% for the VLA at K-band, and $\sim10\%$ for the GBT\footnote{\url{https://www.gb.nrao.edu/GBTCAL/15A486/}}, this value, within the errors, is consistent with that measured by \cite{Braatz_2008} in NGC~1106.
The flux density of the water maser from our observations, along with the continuum flux densities, is given in Table \ref{tab:flux}.

\begin{deluxetable*}{ccccc}[htpb!]
\tablecaption{VLA A-configuration Line and Continuum Flux Density Measurements for NGC~1106 \label{tab:flux}}
\tablecolumns{4}
\tablehead{ \multicolumn{1}{c}{Source} & \multicolumn{1}{c}{Right Ascension} & \multicolumn{1}{c}{Declination} & \multicolumn{1}{c}{Flux Density $\pm$ Error} \\ & J2000 & J2000 & mJy}   
\startdata
\hline
Water Maser             & 2$^{\text{h}}$ 50 $^{\text{m}}$ 40\fs263    & 41\arcdeg 40\arcmin 15\fs90   & $1.02 \pm 0.18$ \\
\hline
Southwest Lobe          & 2$^{\text{h}}$ 50 $^{\text{m}}$ 40\fs478    & 41\arcdeg 40\arcmin 17\fs548  & $2.78 \pm 0.42$  \\
Core                    & 2$^{\text{h}}$ 50 $^{\text{m}}$ 40\fs490    & 41\arcdeg 40\arcmin 17\fs635  & $2.47 \pm 0.4 $ \\
Northeast Lobe          & 2$^{\text{h}}$ 50 $^{\text{m}}$ 40\fs502    & 41\arcdeg 40\arcmin 17\fs805  & $4.51 \pm 0.68$  \\
Extension               & 2$^{\text{h}}$ 50 $^{\text{m}}$ 40\fs473    & 41\arcdeg 40\arcmin 17\fs11   & $0.8 \pm 0.23$  \\
\enddata 
\centering
\end{deluxetable*}

\section{Results}
\label{sec:results}

\begin{figure*}[ht!]
\centering
      \subfloat{{\includegraphics[width=0.45\textwidth]{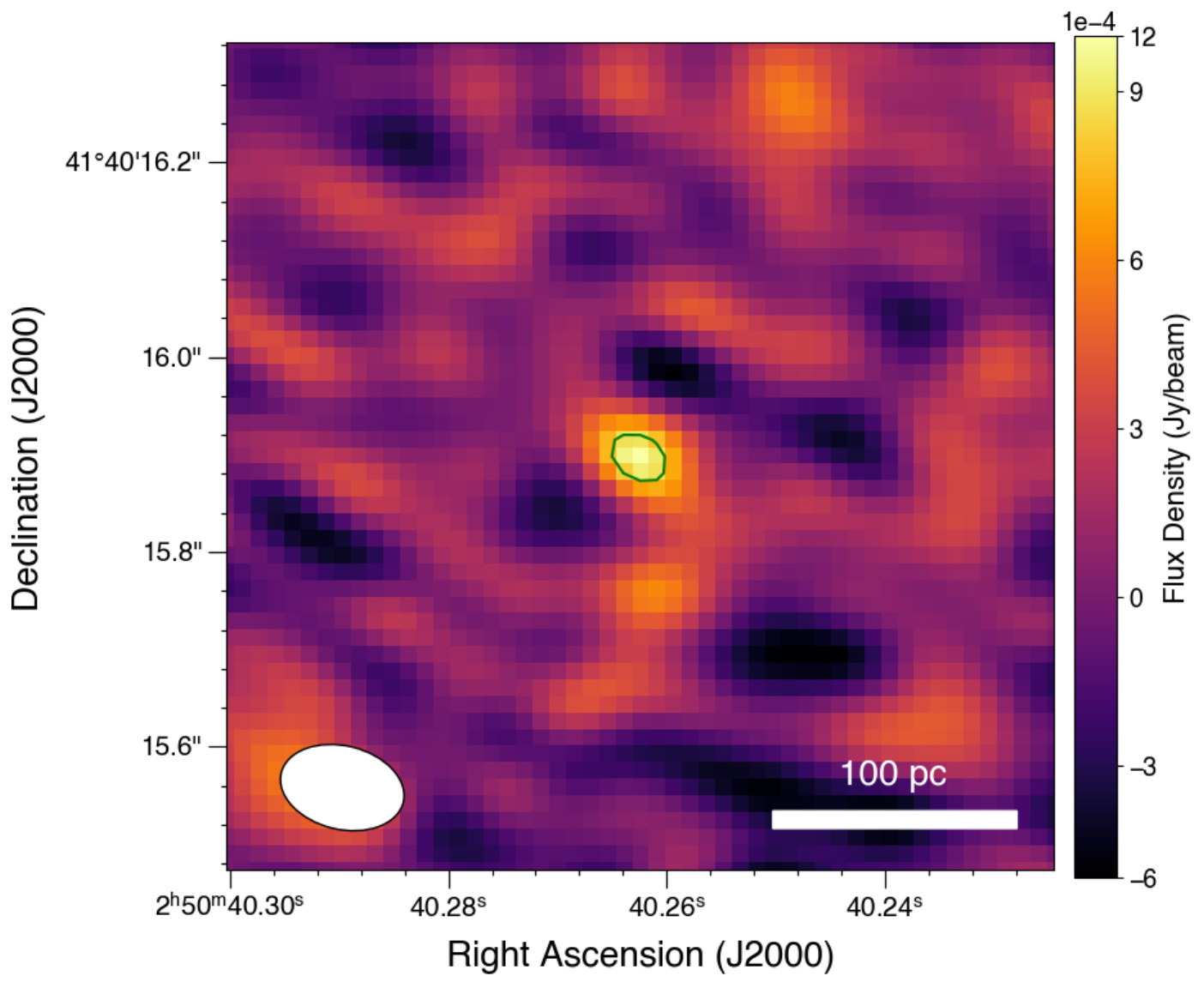} }}%
      \qquad
      \subfloat{{\includegraphics[width=0.5\textwidth]{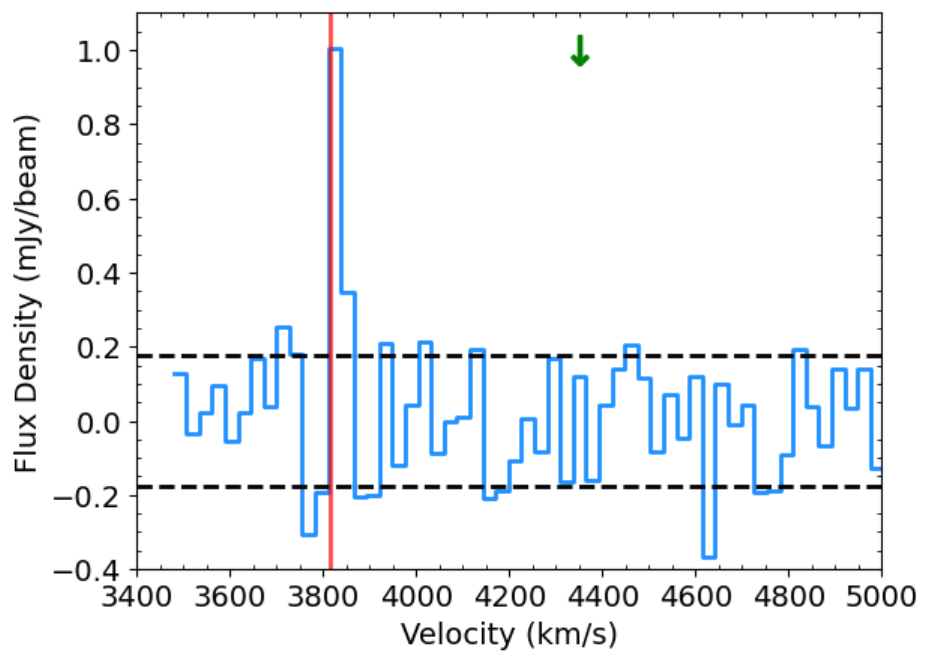} }}%
      \caption{Left: VLA image of the compact water maser in NGC~1106, highlighted by the $5\sigma$ green contour. Right: VLA spectrum of NGC~1106 extracted from the brightest pixel of the water maser with a spectral resolution of $\sim 27$ km s$^{-1}$. The velocity is in the barycentric frame using the optical velocity definition. The water maser is unresolved and located at $3810 \pm 14$ km s$^{-1}$ with a flux density of $1.02 \pm 0.18$\,mJy ($5.8\sigma$ detection), which is shifted by $541 \pm 19$ km s$^{-1}$ with respect to the systemic velocity of host galaxy, denoted with a green arrow \citep[$4351 \pm 13$ km s$^{-1}$;][]{Koss_2022c}. The black dashed lines are the $\pm 1\sigma$ RMS noise of the image cube. The solid vertical red line represents the velocity of the water maser detected by \cite{Braatz_2008} at $3814$ km s$^{-1}$ with the GBT. }
      \label{fig:water_maser_spectra}
\end{figure*}

Our observation of the water maser in NGC~1106 with the VLA in A-configuration yields a high angular resolution of $0$\farcs$131$ but a lower spectral resolution than those in the GBT and DSN observations (see Table \ref{tab:vla_obs}). \cite{Braatz_2008} detected a single narrow ($1$ km s$^{-1}$) water maser feature at a velocity of $3814$ km s$^{-1}$. This is offset from the systematic velocity by $537$ km s$^{-1}$~\footnote{\cite{Braatz_2008} originally reported $4337$ km s$^{-1}$ for the systematic velocity of NGC~1106. We are using the updated velocity of $4351 \pm 13$ km s$^{-1}$ for NGC~1106 \citep{Koss_2022c}}. Our VLA A-configuration observations tentatively detect ($5.8\sigma$) a water maser at $3810 \pm 14$ km s$^{-1}$, which is offset from the systematic velocity by $541\pm30$ km s$^{-1}$. This can be seen in Figure \ref{fig:water_maser_spectra}. 

\begin{figure*}[ht!]
  \begin{subfigure}[b]{0.45\textwidth}
    \includegraphics[width=\textwidth]{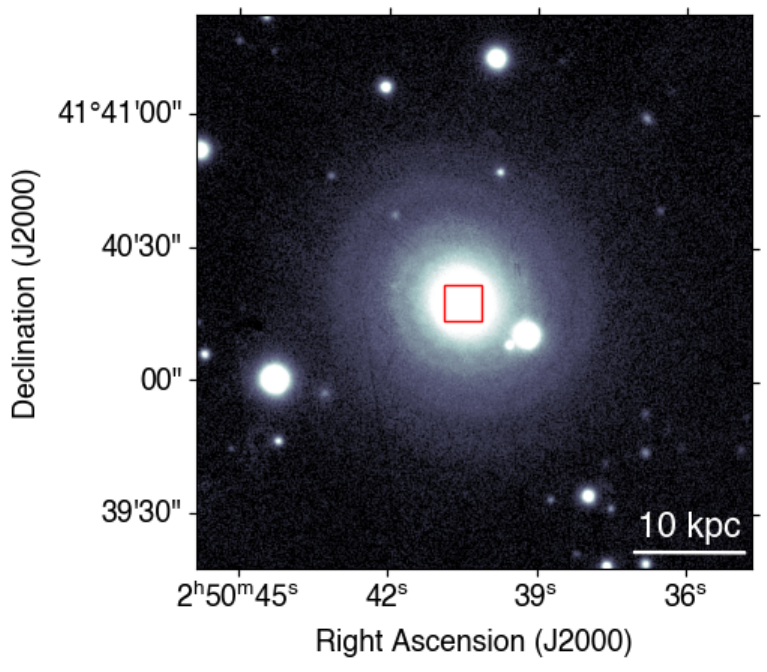}
  \end{subfigure}
  \hfill
  \begin{subfigure}[b]{0.55\textwidth}
    \includegraphics[width=\textwidth]{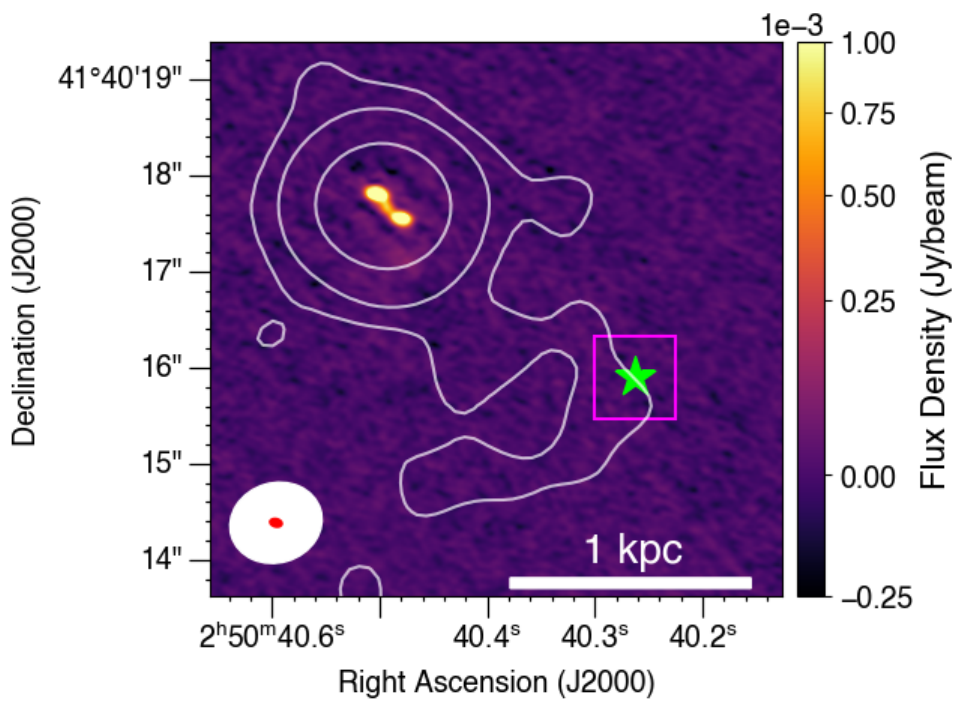}
  \end{subfigure}
  \caption{Left: PanSTARRS g-band image of NGC~1106 \citep{Chambers_2016, Flewelling_2020, Magnier_2020a, Magnier_2020b, Magnier_2020c, Waters_2020}. The red box represents the region probed with the VLA A-configuration and C-configuration observations shown on the right \citep{Smith_2016, Smith_2020b, 2025_Magno}. Right: VLA A-configuration 22 GHz continuum image with 0.13\arcsec resolution, in color scale, showing the jet-like structure present in the core of NGC~1106. The white contours are the VLA 22 GHz C-configuration observation at 1\arcsec resolution \citep{Smith_2016, Smith_2020b, 2025_Magno}. The synthesized beams for the A- and C-configurations are shown in red and white, respectively. The magenta square represents the field-of-view of the water maser image shown in Figure~\ref{fig:water_maser_spectra}. The jet-like structure's axis, as seen in color, is aligned with the larger, $1$\arcsec \, extension. The water maser, denoted by the green star, is located on the boundary of the $1$\arcsec \, contour. }
  \label{fig:radio_outflow_image}
\end{figure*}

The water maser emission in NGC~1106 is clearly not located within the unresolved nuclear core, and there is no $\ge 5\sigma$ detection at all within 0.5\arcsec of the nucleus; the only detection is $\sim1$ kpc away from the core. The VLA results show that the water maser feature is located not in the accretion disk, but $3.12$\arcsec \, or $930$ parsecs from the nuclear region of NGC 1106. Our observing setup included line-free spectral windows, allowing us to image the continuum emission at a resolution of 0\arcsec.13. This revealed an elongated radio structure within NGC~1106 and a faint radio extension south of it. For simplicity, we will refer to this as a jet-like structure, but further follow-up is needed to verify its origin. Comparing our jet-like structure to a 22 GHz $1$\arcsec \, resolution continuum image \citep{Smith_2016, Smith_2020b, 2025_Magno}, we can see that the jet-like structure aligns with the 1\arcsec \, resolution image, which has a compact core along with a radio spur that extends $3$\arcsec \, or $\sim1$ kpc. Most importantly, we can see that the water maser feature in NGC~1106 is on the boundary of the extended continuum emission of the $1$\arcsec \, image (Figure \ref{fig:radio_outflow_image}). Flux densities extracted from our observation of the jet-like structure and the radio extension are given in Table \ref{tab:flux}. 

Due to the alignment of the extended radio emission in the $1$\arcsec\, observations, the jet-like structure axis, and the water maser position, we suggest that the maser arises from AGN feedback occurring as a result of the interaction between the AGN-driven outflow and the ISM. Additionally, the high-velocity ($541$ km s$^{-1}$) offset of the water maser feature may indicate a shock plowing through the ISM \citep{Allen_2008, Cazzoli_2022, Fischer_2025}. We did not see any water masers arising from the nucleus of NGC~1106, but this could be a result of the high noise in our image cube ($185$ $\mu$Jy beam$^{-1}$), or perhaps there is significant variability in the intensity of the water masers \citep[e.g.,][]{Maloney_2002b, Braatz_2004, Tarchi_2005, Tarchi_2007a}. 

\section{Discussion} \label{sec:discussion}
The tentative discovery of a resolved water maser feature in an AGN along the jet/outflow axis both provides an important new observable for feedback energetics and warns against using water maser velocities from single-dish observations to measure black hole mass. It is important, therefore, to explore other possible origins of such a maser signal and to put it into proper context with other extragalactic water masers.

\subsection{Black Hole Mass Estimates}

Extranuclear water maser features have implications for maser derived black hole masses. In fact, \cite{Gao_2017} states that for Mrk~1029 and NGC~1320, the velocity structures of the water maser lines are sufficiently complex to prevent robust black hole mass measurements. It could be possible that these AGN have shock-driven water masers in addition to disk water masers, which would complicate the picture immensely. In fact, \cite{Xanthopoulos_2001} conducted a VLBI study of Mrk~348 and unveiled a small-scale jet that powers the water maser via shocks. All of this suggests that water maser emission may not always trace circumnuclear rotation in the accretion disk, but in some cases could instead arise in shocked gas associated with AGN feedback interacting with the host material, further emphasizing the need for a multi-wavelength approach to interpreting these systems.

If a high-velocity water maser feature was assumed to be at $\sim 0.1$ pc in the accretion disk but, in reality, it was $\sim1$ kpc away, as seen here in NGC~1106, then the calculated enclosed mass would be drastically miscalculated. To calculate how much the enclosed mass is off by, we assume Keplerian rotation \citep[e.g.,][]{Kuo_2018}:
\begin{equation}
    v_{\text{orb}} = \sqrt{\frac{G M_{\text{BH}}}{r}}
\end{equation}
where r is the orbital radius of a particular maser spot, $v_{orb}$ is the velocity of the water maser, $G$ is the gravitational constant, and $M_{\text{BH}}$ is the black hole mass. If a water maser feature is assumed to be at 0.1 pc, but is actually at 1 kpc, then the inferred enclosed mass is off by $\sim 10^{4}$ M$_{\odot}$. In any case, in a shock-driven scenario the velocity is not orbital, so the enclosed mass calculation has no meaning. Isolated high-velocity features or double-peaked spectral lines without a well-detected systemic component are the most vulnerable to contamination by outflow-driven masers. This is consistent with what \cite{Kuo_2020} finds: double-peaked megamaser systems are dynamically perturbed in ways suggestive of AGN wind or jet impact, and NGC 1320, in particular, lacks systemic maser features while showing counter-rotating gas at 300 pc scales. Our result extends this picture, showing that AGN-driven shocks can produce maser emission not only at sub-pc scales within the perturbed disk but also at kiloparsec scales at the outflow–ISM interface, far outside any disk geometry. Observational diagnostics that flag potential contamination would be the absence or weakness of systemic maser features, alignment of high-velocity features with known radio or ionized-outflow axes, and velocity offsets that are comparable to or exceed independently estimated outflow velocities from X-ray or optical tracers. Therefore, when moving from single-dish measurements to higher-resolution observations, it is necessary to search for signs of AGN outflow activity, such as the presence of outflow and/or high-velocity offset lines without a Keplerian signature.

\subsection{Isotropic Luminosity} \label{sec:isotropic}
By calculating the isotropic luminosity, one can try to distinguish the origin of the water maser, although luminosity alone may not uniquely determine the powering mechanism. Water masers with isotropic luminosities of $L_{\text{H}_2\text{O}} > 20 \, L_{\odot}$ are traditionally classified as megamasers and are frequently associated with AGN activity, while in star-forming regions they typically have an isotropic luminosity of $L_{\text{H}_2\text{O}} < 0.1 \, L_{\odot}$. The intermediate range between these two regimes consists of kilomasers, which are primarily thought to originate from supernova remnants or ultradense star-forming regions \citep{Tarchi_2002, Castangia_2008, Darling_2008, Darling_2016}. 

To calculate the isotropic line luminosity, we will use the following equation:
\begin{equation}
    L_{\text{iso}} = \frac{23.1 \, L_{\odot} \times S_{\text{int}} \times D_L^2}{1+z}
\end{equation}
where $S_{\text{int}}$ is the integrated line flux density in mJy km s$^{-1}$2, $D_L$ is the luminosity distance in gigaparsecs, and $z$ is the cosmological redshift \citep{Darling_2017}. For reference, \cite{Braatz_2008} found a isotropic luminosity of 8 L$_{\odot}$ for this water maser using the GBT with a higher spectral resolution than what we have with the VLA. For this calculation, we will use the flux density from the tentative detection with the VLA ($1.02\pm0.18$ mJy), and we arrive at a isotropic luminosity of $2.5$ L$_{\odot}$. It should be noted that this isotropic luminosity should be treated as a lower limit due to our lower spectral resolution. With this information, we can classify the water maser in NGC~1106 as a kilomaser. It is entirely possible that the water maser within NGC~1106 originates from a ultra-dense star-forming region or supernovae remnant, but we need further multi-wavelength follow-up to verify this similar to what \cite{Tarchi_2011} did for NGC 3356. If this ends up being true, then NGC~1106 could be a new addition to the few galaxies in the local universe that have evidence for positive AGN feedback based off the information presented in Figure \ref{fig:radio_outflow_image}.

\subsection{Water Masers from Star Formation}
It is possible that the off-nuclear water maser in NGC~1106 originates from an intense star-forming region coincident with the $1$\arcsec \, extension. A useful comparison is W3(OH), one of the brightest Galactic water maser sources \citep{Dreher_1981, Dreher_1984, Yamada_2024}. W3(OH) is a high-mass, ultra-dense star-forming region in the W3 complex where stellar shocks from stellar winds heat the dense gas, producing strong water maser emission \citep{Dreher_1981, Dreher_1984, Feigelson_2008}. Throughout this section, we will compare both the GBT flux density \citep{Braatz_2008} and our VLA flux density to what has been observed in W3(OH). 

W3(OH) is located within the Milky Way at a distance of $2.04\pm0.07$ kpc from Earth \citep{Hachisuka_2006}. Considering that NGC~1106 is at a distance of $63.2$ Mpc \citep[$z = 0.014513 \pm 4.34\times10^{-5}$;][]{Koss_2022b}, we find that if a similar water maser emitting region were to occur within the Milky Way, it would have a flux density of $1.02 \times 10^6$ Jy. When we use the reported GBT value of 74 mJy \citep{Braatz_2008}, we obtain $6.89\times10^{7}$ Jy. For reference, the water maser-emitting regions within W3(OH) can range from mJy to a few hundred Jy \citep{Cohen_1979, Hachisuka_2006}. Even at its brightest, W3(OH) is several orders of magnitude fainter than the maser observed in NGC~1106 if they were both at the same distance, making it unlikely that the NGC~1106 maser arises from a star-forming region like W3(OH). 

\subsection{Comparison to Other Extragalactic Water Masers} 

The merger system NGC~3256 has two off-nuclear water masers, each located $\sim1$\arcsec \, away from one of the two radio continuum cores \citep{Surcis_2009}. A comparison of the water maser velocities and the $^{12}$CO($2-1)$ velocities revealed discrepancies on the order of $25$ km s$^{-1}$. From this, \cite{Surcis_2009} determined that the water masers in NGC~3256 originated in two hot spots associated with star formation activity and not the AGN. The typical velocity offset for these water masers is on the order of $20-40$ km s$^{-1}$ with respect to the systemic velocity. This is considerably lower than the velocity offset for NGC~1106, which is $541$ km s$^{-1}$. 

NGC~3556 has nuclear water maser activity that was observed with the VLA in CnB and DnA configurations $\sim4$ years apart \citep{Tarchi_2011}. Its nature was determined through a multi-wavelength analysis in which the lack of typical AGN signatures pointed toward star formation. In fact, a compact source detected at 1.4 GHz appears to be a group of supernova remnants following the HI gas disk \citep{Irwin_2000, Tarchi_2011}.

Centaurus A is a radio-loud galaxy and is known to host a nuclear water maser. This kilomaser was first detected with the Australia Telescope Compact Array (ATCA) over the course of $\sim2$ years \citep{Ott_2013}. The water maser line in Centaurus A are offset by $415$ km s$^{-1}$ and they have a width of $\sim 120$ km s$^{-1}$. Based off this information and the fact that the line profile is fairly smooth, \cite{Ott_2013} concludes that this water maser is jet-driven, and it is located $\leq 3$ pc away from the central SMBH.

NGC 1052 is a known jetted LINER Seyfert 2 galaxy that also hosts a nuclear 22 GHz water megamaser \citep{Braatz_1994, Braatz_1997, Kameno_2005}. Typically, this AGN has a broad maser line profile (FWHM $> 100$ km s$^{-1}$ with an isotropic luminosity of 125 L$_{\odot}$ \citep{Braatz_1994, Braatz_1997}. \cite{Kameno_2005} observed NGC 1052 with the Nobeyama 45\,m radio telescope, and they uncovered a new narrow (FWHM of 21 km s$^{-1}$) water maser feature that is offset by 328 km s$^{-1}$ with respect to the systemic velocity. Since this feature was discovered with a single dish telescope, \cite{Kameno_2005} was not able to spatially localize the narrow water maser feature, but they speculated that due to the high-velocity offset, it likely arose from the inner 0.05 pc of the central engine. Recently, \cite{Kameno_2024} detected the 321 GHz water maser transition in NGC 1052 using ALMA long-baseline observations, partially resolving the emission at the jet-torus boundary within the inner parsec, and adding further credence to the origin of the 22 GHz narrow high-velocity water maser feature \cite{Kameno_2005} observed.

NGC 1068 is a prototypical Seyfert 2 galaxy that hosts an AGN driven wind along with nuclear and off-nuclear water masers \citep[e.g.,][]{Gallimore_1996b, Braatz_1997, Fischer_2023, Hagiwara_2024}. \cite{Gallimore_1996b} used the VLA A-configuration and found an off-nuclear water maser component in this galaxy, which is located 0.\arcsec3 ($\sim20$ pc) away from the core with a velocity offset of $\sim140$ km s$^{-1}$,a width of $\sim50$ km s$^{-1}$ and an isotropic luminosity of $\sim26$ L$_{\odot}$. These water masers occur at the shock interface with a molecular cloud \citep{Gallimore_1996a, Gallimore_1996b, Fischer_2023, Hagiwara_2024}. \cite{Hagiwara_2024} observed this off-nuclear component with the Effelsberg 100-m Radio Telescope, and used MERLIN to localize the water maser. Then they followed up with ALMA to map the molecular gas around the offset water maser, and found that the off-nuclear masers nearly overlap with a ring-like molecular gas structure. These water masers are tracing an evolving shocked structure likely coming from the interaction of the radio outflow and circumnuclear medium. From their MERLIN observations, they measured an isotropic luminosity of $\sim0.1$ L$_{\odot}$ for the off-nuclear maser component. Due to the change in the spectral profile, it is possible that the maser has faded between \cite{Gallimore_1996b}'s initial observation with the VLA and \cite{Hagiwara_2024}'s observation with MERLIN since these observations occurred roughly 20 years apart. Additionally, \cite{Hagiwara_2024} found that the off-nuclear water maser central velocity has shifted towards the blue over the course of 35 years.

The above demonstrates that extragalactic water masers with a star-forming origin tend to be similar in some respects (velocity dynamics and brightness temperature) to those observed in W3(OH), while water masers with a jet-driven origin have a larger velocity offset with respect to the host systemic velocity. NGC~1106 stands out from these in three ways: 1. The water maser is on the boundary of a radio extension that is along the axis of AGN feedback. 2. The velocity of the water maser is shifted by $541$ km s$^{-1}$ from the systemic velocity of NGC 1106, substantially larger than the velocity offsets typically observed in star-formation-associated water masers and more consistent with the high-velocity components commonly observed in AGN-associated maser systems. For example, the star-forming water masers in NGC 3256 are offset from the host galaxy systemic velocity by only 20–40 km s$^{-1}$ \citep{Surcis_2009}, while the high-velocity features in NGC 1068, Centarus A, and NGC 1052 are offset by 300 km s$^{-1}$ or more \citep{Gallimore_1996b, Kameno_2005, Ott_2013, Hagiwara_2024}. The large velocity offset observed in NGC 1106 suggests that the water maser is unlikely to be associated with a typical star-forming region, and is more likely to be shock-driven. 3. The width of the water maser is very narrow \citep[1 km s$^{-1}$;][]{Braatz_2008}, which does not agree with the line width of the jet-driven water maser in Centaurus A \citep{Ott_2013}. The narrow high-velocity features seen in NGC 1052 and NGC 1068 are much smaller than what was observed in Centaurus A, but they are still an order of magnitude larger than what has been observed in NGC 1106. This is likely due to the distances at which the shock-front has impacted a molecular cloud. For NGC 1106, the shock is occurring 1 kpc away from the AGN, while, in NGC 1068 the shock-front occurs $\sim20$ pc away from the AGN. Therefore, it is entirely possible that the water maser in NGC 1106 is likely induced by an AGN outflow. A future variability study of the water maser in NGC 1106 could provide further evidence for an AGN-driven origin by testing for changes in its isotropic luminosity expected from an AGN outflow propagating through the ISM \citep{Hagiwara_2024}.

\section{Conclusions} 
\label{sec:conclusions}
We have conducted a 22 GHz VLA A-configuration 22 GHz study on the water maser emission in the \textit{Swift}-BAT AGN, NGC~1106. We have a tentative detection of the water kilomaser feature with the VLA at $3763 \pm 14$ km s$^{-1}$, offset by 541 km s$^{-1}$ from the systematic velocity. We were also able to localize the water maser-emitting region to a small, unresolved source $930$ parsecs away from the radio core, directly along the axis of the radio structure. Thanks to a previous VLA C-configuration observation \citep{Smith_2016, Smith_2020b, 2025_Magno}, we identified the water maser-emitting region as coincident with the edge of a radio extension with the same position angle as the radio structure in our VLA A-configuration continuum image. A comparison to different sources of water masers (star-formation and AGN driven) suggests that the NGC~1106 maser is likely to be due to an AGN driven outflow. To definitively determine the origin of this water maser, we need higher spatial and spectral resolution along with a slew of multi-wavelength observations.

We conclude by remarking again on the implications of extranuclear masers driven by shocks for maser-derived black hole mass measurements. Most maser spectra are taken by single-dish telescopes with poor spatial resolution. Interpretation of complex maser velocity structures as inherent to a rotating accretion disk will lead to drastic miscalculations of black hole masses if shock-driven velocity components in maser lines are common.

\begin{acknowledgments}
The authors are very thankful for the referee's valuable comments, which helped shape this manuscript. MK acknowledges support from NASA through ADAP award 80NSSC22K1126.  This research has made use of the SIMBAD database, CDS, Strasbourg Astronomical Observatory, France \citep{Simbad}. The Pan-STARRS1 Surveys (PS1) and the PS1 public science archive have been made possible through contributions by the Institute for Astronomy, the University of Hawaii, the Pan-STARRS Project Office, the Max-Planck Society and its participating institutes, the Max Planck Institute for Astronomy, Heidelberg and the Max Planck Institute for Extraterrestrial Physics, Garching, The Johns Hopkins University, Durham University, the University of Edinburgh, the Queen's University Belfast, the Harvard-Smithsonian Center for Astrophysics, the Las Cumbres Observatory Global Telescope Network Incorporated, the National Central University of Taiwan, the Space Telescope Science Institute, the National Aeronautics and Space Administration under Grant No. NNX08AR22G issued through the Planetary Science Division of the NASA Science Mission Directorate, the National Science Foundation Grant No. AST–1238877, the University of Maryland, Eotvos Lorand University (ELTE), the Los Alamos National Laboratory, and the Gordon and Betty Moore Foundation. 
\end{acknowledgments}





%
\facilities{VLA}

\software{CASA \citep{McMullin_2007, CASA_2020a, CASA_2020b, CASA_2022a, CASA_2022b, TheCASAteam_2022}, CARTA \citep{comrie_2024}, AIPS \citep{Greisen_2003}, astropy \citep{2013A&A...558A..33A, 2018AJ....156..123A, 2022ApJ...935..167A}, APLpy \citep{aplpy2012, aplpy2019}, numpy \citep{harris2020array}, Matplotlib \citep{Hunter:2007}, spectral-cube \citep{2019zndo2573901G}, scipy \citep{scipy}, pandas \citep{pandas1, pandas2}         }


\bibliography{reference}{}
\bibliographystyle{aasjournalv7}



\end{document}